\documentclass[aps,twocolumn,showpacs,preprintnumbers,amsmath,amssymb,prl]{revtex4-1}
\usepackage{epsf,amsmath,amssymb,verbatim,color,multirow,pifont}
\usepackage{graphicx}

\begin{document}
\title{Two opposite hysteresis curves in semiconductors with mobile dopants}
\author{J.S.~Lee$^{1,2}$, S.B.~Lee$^1$}
\author{B.~Kahng$^{1}$} \email{bkahng@snu.ac.kr}
\author{T.W.~Noh$^{1}$} \email{twnoh@snu.ac.kr}
\affiliation{{$^1$Department of Physics
and Astronomy, Seoul National University, Seoul 151-747, Korea}\\
{$^2$School of Physics, Korea Institute for
Advanced Study, Seoul 130-722, Republic of Korea}}

\date{\today}

\begin{abstract}
Recent experimental researches on semiconductors with mobile dopants (SMD) have reported unconventional hysteretic current-voltage ($I$-$V$) curves, which form dynamically in either one of the two opposite directions, the counter-figure-eight and figure-eight ways. However the fundamental theory for the formation of the two directions is still absent, and this poses a major barrier for researches oriented to applications. Here, we introduce a theoretical model to explain the origin of the two directions, and find that the two ways originate from the spatial inhomogeneity of the dopant distribution in SMD. The counter-figure-eight (figure-eight) way of the hysteretic curve is obtained when dopants are driven from far from (near) the Schottky interface to the opposite side in the SMD. This finding indicates that the directions of hysteretic curve in SMD can be controlled.
\end{abstract}


\maketitle
\begin{figure}[t]
\includegraphics[width=1.0\linewidth]{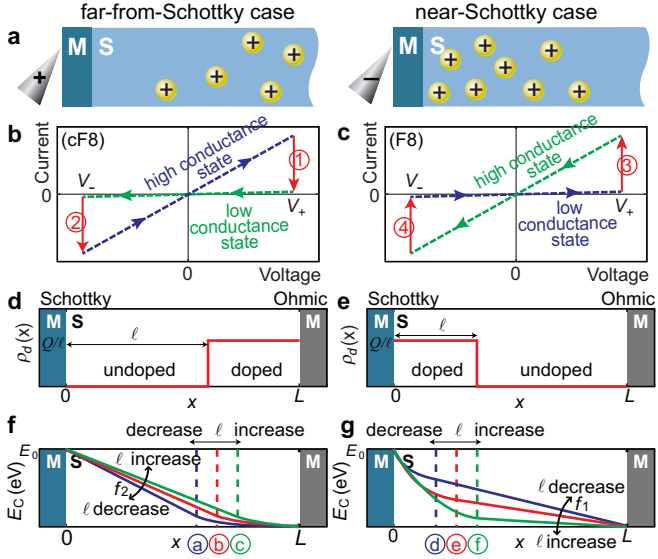}
\caption{(Color online) (a) Diagrams of a $\textrm{O}_\textrm{v}$-based SMD. Dopants can be repelled or attracted by applying a positive or negative bias, respectively. (b) and (c) are the schematics for counter-Figure-8 and Figure-8 $I$-$V$ hysteresis curves, respectively. (d)-(g) one-dimensional SMD model. (d) and (e) show the dopant density distribution $\rho_d (x)$ for the far-from-Schottky and near-Schottky cases, respectively. (f) and (g) show the effects of donor movement on the Schottky barrier for the far-from-Schottky and near-Schottky cases, respectively.}\label{analytic}
\end{figure}

The successful application of semiconductor devices over a wide range of electronic devices is attributable to their unique electrical properties, which are sensitive to the configuration of their dopants. Generally, dopants are assumed to be immobile. What happens if the dopants are mobile? Recent research on semiconductors with mobile dopants (SMD) such as oxygen vacancies ($\textrm{O}_\textrm{v}$)~\cite{Waser2007,Yang2009} indicates that SMD demonstrates its conductance change due to the alteration of the dopant distribution in SMD~\cite{HPnature2008}. As an example, consider a metal(M)-semiconductor(S) Schottky contact (n-type) as shown in Fig.~\ref{analytic}(a). When a positive (negative) external bias $V_\textrm{ext}$ is applied to the metal, the dopants are pushed away from (attracted toward) the metal, as a result, the dopant distribution is non-uniformly altered. Because the interfacial band structure of a semiconductor is sensitive to the concentration of the dopants, different dopant distribution causes different conductance state of the semiconductor. This conductance change is applicable to various fields of science and engineering such as resistive switching phenoemena~\cite{Waser2009R}, memristive system~\cite{HPnature2008}, neuroscience~\cite{neuro1}, and non-volatile memory devices~\cite{Yang2008,Janousch,Nian,MJLee2011}, and thus it has received great interests.

Such conductance change results in a hysteretic current-voltage ($I$-$V$) curve. One conventional explanation for the hysteretic curve is as follows~\cite{Yang2009}. When a positive (negative) bias $V_+$ ($V_-$) is applied, the donor concentration $n_\textrm{d}$ becomes low (high) near the Schottky interface. Then the Schottky barrier width $w_\textrm{sb}$ increases (descreases) because $w_\textrm{sb} \propto 1/\sqrt{n_\textrm{d}}$~\cite{Sze}, thus, the conductance decreases (increases) as denoted by \textcircled{\footnotesize 1} (\textcircled{\footnotesize 2}) in Fig.~\ref{analytic}(b). This directional hysteretic curve is called the counter-figure-eight (cF8) curve~\cite{Shibuya} and has been found in many materials~\cite{Janousch,MJLee2011,Szot}.  

However, the opposite way of curve (Fig.~\ref{analytic}(c)), called the figure-eight (F8) curve, has also been reported in many literatures~\cite{Yang2008,Sawa,Sawa2}. Moreover, the coexistence of both directions in a single sample has been reported~\cite{Yang2009,Muen,Shibuya}. To understand the origin of the two directions, several experiments and heuristic arguments have been presented. For example, Yang et al.~\cite{Yang2009} suggested that the cF8 and F8 curves are derived from the top and bottom Schottky interfaces, respectively. However, Shibuya et al.~\cite{Shibuya} hypothesized that the cF8 curve arises from $\textrm{O}_\textrm{v}$ movement through conducting filaments inside the sample, whereas the F8 curve has a purely electronic origin. Subsequently, the same authors suggested that the cF8 and F8 curves originate from the respective inhomogeneous (or filamentary) and homogeneous distributions of $\textrm{O}_\textrm{v}$ parallel to the interface \cite{Muen}. 

Despite these experimental results and heuristic arguments, the origin of the two ways of hysteretic $I$-$V$ curves has not been elucidated theoretically yet. In this Letter, we theoretically demonstrate that the two ways of $I$-$V$ hysteretic curves intrinsically appear in the SMD, resulting from the spatial inhomogeneous distribution of dopants. When most dopants are located far from (near) the Schottky interface, the cF8 (F8) curve appears.  

We first introduce a simple theoretical model in one dimension to illustrate the mechanism of the two ways in SMD depending on the initial dopant distribution. In this model, the semiconductor is in contact with the metals located at $x=0$ and $x=L$ to form the Schottky and Ohmic interfaces (Fig.~\ref{analytic}(d)) with the boundary conditions $E_C (x=0)=E_0$ and $E_C (x=L)=0$, respectively. We consider two different cases of initial dopant density distributions: dopants are located 1) far from (Fig.~\ref{analytic} (d)) and 2) near (Fig.~\ref{analytic} (e)) the Schottky interface.
For simplicity, we assume that the dopant density distribution $\rho_d (x)$ is constant in space. Then, for doped region, $\rho_d=Q/(L-\ell)$ in the region [$x=\ell$, $L$] for the far-from-Schottky case and $\rho_d=Q/\ell$ in [$x=0$, $\ell$] for the near-Schottky case, where $Q$ is the total amount of dopants in a semiconductor and assumed to be a conserved quantity. For undoped region, $\rho_d=0$. This simplification is very useful to capture the essential mechanism of the two ways of the hyteresis curves.  
We assume that the electrons are fully depleted in the doped region for analytic calculation. Non-constant $\rho_d (x)$ case and not-fully depleted cases will be treated numerically later. Under this simplified circumstance, the position-dependent conduction band $E_C(x)$ 
can then be calculated by solving the Poisson's equation \cite{Neamen}, $\nabla^2 E_C (x)=e\rho_{sc} (x)/\epsilon$, where $e$ is the electronic charge, $\rho_{sc} (x)$ is the space charge density, and $\epsilon$ is the permittivity of the semiconductor. Note that $\rho_{sc} (x)=q\rho_d(x)$, where $q$ is the dopant charge. Here, we deal with the case $q>0$ (n-type semiconductor).

We first consider the far-from-Schottky case. The Poisson's equations for $E_C (x)$ in the regions $x< \ell$ and $x > \ell $ become $d^2 E_C (x)/dx^2 =0$ and $d^2 E_C (x)/dx^2 =qeQ/\epsilon (L-\ell)$, respectively. Using the boundary conditions, $E_C(x=0)=E_0$ and $E_C(L)=0$, and continuity at $x=\ell$, we can easily obtain $E_C(x)$ in the whole range. Particularly for $x< \ell$, we obtain that 
\begin{equation}
E_C (x)=f_1 x+E_0,~~\textrm{where}~f_1=-\frac{E_0}{L}-\frac{qeQ(L-\ell)}{2\epsilon L}. \label{f1}
\end{equation}
Here, $f_1$ is the slope of $E_C$ in the undopped region. 
If $\ell$ is initially located at \textcircled{\footnotesize a} in Fig.~\ref{analytic}(f) and a positive bias $V_+$ is applied, $\ell$ increases as the direction \textcircled{\footnotesize a}$\rightarrow$\textcircled{\footnotesize b}$\rightarrow$\textcircled{\footnotesize c}. Then $f_1$ increases or the slope in the undopped region becomes less steeper (Eq.~(\ref{f1})) as shown in Fig.~\ref{analytic}(f), which makes the Schottky barrier width $w_\textrm{sb}$ thicker. Therefore, the conductance decreases, which corresponds to the conductance change denoted by \textcircled{\footnotesize 1} in Fig.~\ref{analytic}(b). If a negative bias $V_-$ is applied to this low conductance state, $\ell$ will change reversely as \textcircled{\footnotesize c}$\rightarrow$\textcircled{\footnotesize b}$\rightarrow$\textcircled{\footnotesize a}. Then the conductance increases as denoted by \textcircled{\footnotesize 2} in Fig.~\ref{analytic}(b). This result agrees with the conventional explanation for cF8 curve.

For the near-Schottky case, the calculation for $E_C(x)$ can be performed similarly. The Poisson's equations for $x< \ell$ and $x> \ell$ become $d^2 E_C (x)/dx^2 =qeQ/\epsilon \ell$ and $d^2 E_C (x)/dx^2 =0$, respectively. 
For $x>\ell$, we obtain that 
\begin{equation}
E_C (x)=f_2 (x-L),~~\textrm{where}~f_2=-\frac{E_0}{L}+\frac{qeQ\ell}{2\epsilon L}. \label{f2}
\end{equation}
If initial $\ell$ is located at \textcircled{\footnotesize d} in Fig.~\ref{analytic}(g), $V_+$ makes $\ell$ increase as the direction \textcircled{\footnotesize d}$\rightarrow$\textcircled{\footnotesize e}$\rightarrow$\textcircled{\footnotesize f}. Then, by the similar explanation as the far-from-Schottky case, $w_\textrm{sb}$ becomes thinner as shown in Fig.~\ref{analytic}(g) and the conductance increases, which corresponds to the conductance change denoted by \textcircled{\footnotesize 3} in Fig.~\ref{analytic}(c). If $V_-$ is applied to this high conductance state, reverse process occurs, which causes the conductance decreases as denoted by \textcircled{\footnotesize 4} in Fig.~\ref{analytic}(c). Therefore, this result verifies that F8 curve intrinsically appears in SMD without the assumption of the electronic function or the two Schottky interfaces.

\begin{figure}[t]
\includegraphics[width=0.9\linewidth]{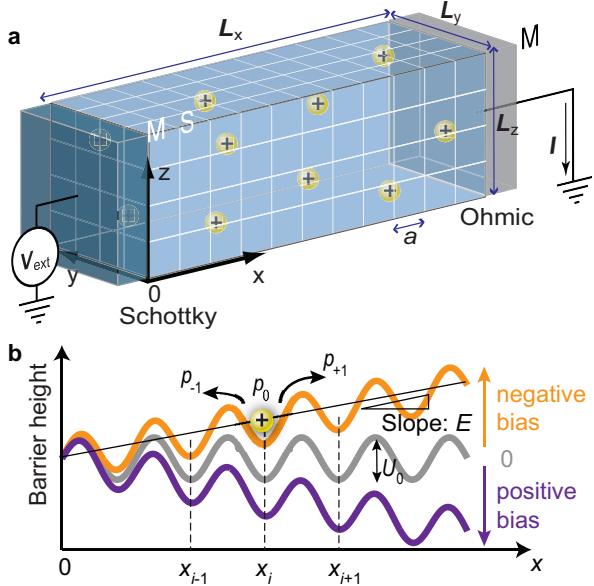}
\caption{(Color online) (a) Configuration of the simulation. The three-dimensional lattice (semiconductor) is in contact with different metals to form a Schottky contact at $x = 0$ and an Ohmic contact at $x = L_x$. Donors are denoted by yellow circles. (b) Periodic potential energy of the donors (grey curve), where local minima correspond to the lattice sites. $U_0$ is the energy barrier height against the movement of a donor. When a negative (positive) voltage is applied, the periodic potential energy increases (decreases), as indicated by the orange (purple) curve. The slope $E$ corresponds to the applied electric field. Donors move according to the hopping probabilities $p_0$, $p_{+1}$, and $p_{-1}$ (Eq.~(\ref{prob})).
}\label{schematic}
\end{figure}

Using numerical simulation, now, we will show that the above analysis is still valid without the assumptions of non-uniform $\rho_d(x)$ and not-fully depleted doped region. For quantitative calculation, we use the parameters for Pt-SrTiO$_3$ contact.
Let us consider a three-dimensional lattice (lattice constant $a=0.39$ nm) whose lengths in $x$-, $y$-, and $z$-directions are $L_x$, $L_y$, and $L_z$, respectively, as shown in Fig.~\ref{schematic}(a). Here, $L_x=L_y=L_z=27.3$ nm for manageable calculation. Two different metals are in contact with the lattice at $x = 0$ and $x = L_x$ forming Schottky and Ohmic contacts, respectively. Donors ($\textrm{O}_\textrm{v}$) were distributed on the lattice depending on $\rho_d (x)$. 
Then the position-dependent conduction band $E_C(x_i,y_j,z_k )$ can be calculated numerically by solving the Poisson's equation, $\nabla^2 E_C (x_i,y_j,z_k )=e\rho_{sc} (x_i,y_j,z_k )/\epsilon$. 
However, the calculation of $E_C$ is not straightforward because $\rho_{sc} (x)\neq q\rho_d (x)$. Therefore, we use the self-consistent relaxation method to obtain $\rho_{sc}$ and $E_C$ simultaneously; we divide $\rho_{sc}$ into two parts: $\rho_{sc} = \rho_+ - \rho_-$, where $\rho_+$ and $\rho_-$ are the densities of positive and negative charges, respectively. When the density of donors is sufficiently high compared with the hole density, $\rho_+\approx q\rho_d (x)$. $\rho_- (x_i,y_j,z_k)$ corresponds to the density of electrons and is determined by the following equation \cite{Sze}: 
$\rho_- (x_i,y_j,z_k )=2 N_c/\sqrt{\pi} \int_0^\infty d\eta ~\eta^{0.5}/(1+\exp⁡ [\eta-\beta \{E_F-E_C (x_i,y_j,z_k ) \} ] ),$ 
where $\beta$ is the inverse temperature and $N_c$ is the effective density of the states in the conduction band. We used $N_c = 2.5\times 10^{19} ~\textrm{cm}^{-3}$ because $N_c \sim 10^{19} ~\textrm{cm}^{-3}$ for many semiconductors~\cite{Neamen}. We also confirmed that the essential feature of the simulation is not changed by variation of $N_c$.
Note that we set $E_F=V_\textrm{ext}$ in the semiconductor and assume that the barrier height at the M-S interface is independent of the dopant density~\cite{Shang}. Thus, we can set up the boundary conditions at $x=0$ (ideal Schottky) and $x = L_x$ (ideal Ohmic) interfaces as $E_C (0,y,z)=0.9$ eV~\cite{Robertson} and $E_C (L_x,y,z)=V_\textrm{ext}$, respectively. Here, we neglect the image-charge effect for the sake of simplicity. Inserting $\rho_+ (x_i,y_j,z_k )$ and $\rho_- (x_i,y_j,z_k)$ into the Poisson's equation, we obtain $E_C (x_i,y_j,z_k)$ and $\rho_{sc} (x_i,y_j,z_k)$ simultaneously. To confirm the validity of this technique, we calculate $E_C (x_i,y_j,z_k)$ for a silicon semiconductor with various doping concentrations. The results are presented in Supplemental Material (SM) 1. The resulting $E_C (x_i,y_j,z_k )$ describes the conduction band which is distorted in the insulating region of the Schottky barrier.

Next, using the obtained $E_C (x_i,y_j,z_k )$, the electric current $I$ of the major
carriers (i.e., electrons) can be estimated with the following formula~\cite{Straton}:
\begin{eqnarray}
I&=&\sum_{j,k} \frac{4e\pi m_e}{\beta h^3} \int_0^\infty dE_x P_{j,k} (E_x) \nonumber\\
&\times&  \ln \left( \frac{1+\exp\left(\beta (\xi-E_x) \right)}{1+\exp\left( \beta (\xi-E_x-V_\textrm{ext}) \right)} \right), \label{current}
\end{eqnarray}
where $m_e$ is the free electron mass, $h$ is Planck's constant, and $\xi=\textrm{max}⁡(E_F-E_C)$. $P_{j,k} (E_x )$ is the transition probability that an electron with $x$-directional energy $E_x$ will tunnel through the Schottky barrier at $y = y_j$ and $z = z_k$. In the discrete lattice, $P_{j,k} (E_x )$ can be written as
$
P_{j,k} (E_x ) \approx \exp \left( -\alpha \sum_i a \sqrt{E_C (x_i,y_j,z_k) - E_x} \right),
$
where the summation index $i$ extends over all cases satisfying $E_C (x_i,y_j,z_k )> E_x$ and $\alpha=1.025~ \textrm{eV}^{-0.5} \textrm{\AA}^{-1}$.

We assume a simple hopping motion along the $x$-direction for the donors under a periodic potential with a barrier height $U_0$, as shown in Fig.~\ref{schematic}(b), because it is widely accepted that $\textrm{O}_\textrm{v}$ must overcome an energy barrier to move to the next lattice site \cite{Zhang}. We also assume that a constant electric field $E=-V_\textrm{ext}/L_x$ is formed throughout the semiconductor when $V_\textrm{ext}$ is applied between two electrodes. The validity of the constant $E$-field approximation is discussed in SM2. Thus, when a negative (positive) $V_\textrm{ext}$ is applied, the periodic potential energy for the donors increases (decreases) with a slope of $E$, as shown in Fig.~\ref{schematic}(b). Then, the heights of the left and right energy barriers, compared to the local minimum, become approximately $U_0-aE/2$ and $U_0+aE/2$, respectively. 
The probability of remaining at the original site $x_i$ ($p_0$) is given by the probability that the donor cannot overcome a lower barrier among the two. So, $p_0=1-\exp(-\beta(U_0-a|E|/2) )$. When $V_\textrm{ext}>0$, the probability of moving to site $x_{i-1}$ ($p_{-1}$) is the half of the probability that the donor overcomes the left or higher barrier (another half of the probability should be counted for moving to the opposite direction). So, $p_{-1}=0.5 \exp(-\beta(U_0+a|E|/2) )$. Then $p_{+1}=1-p_{0}-p_{-1}= \exp(-\beta(U_0-a|E|/2) )-0.5 \exp(-\beta(U_0+a|E|/2) )$.
Similarly, when $V_\textrm{ext}<0$, $p_{+1}=0.5 \exp(-\beta (U_0+a|E|/2) )$ and $p_{-1}=\exp(-\beta (U_0-a|E|/2) )-0.5 \exp(-\beta (U_0+a|E|/2) )$. By combining these two cases, we obtain
\begin{eqnarray}
p_{+1}&=&0.5 e^{-\beta U_0 } \left[e^{\beta |E|a/2}+2 \textrm{sgn}( V_\textrm{ext} )  \sinh⁡(\beta |E|a/2) \right], \nonumber\\
p_{-1} &=& 0.5 e^{-\beta U_0 }\left[e^{\beta |E|a/2}-2\textrm{sgn}(V_\textrm{ext} )  \sinh⁡(\beta |E|a/2) \right] , \nonumber\\
p_0 &=& 1-e^{-\beta(U_0-|E|a/2) },
\label{prob}
\end{eqnarray}
where $\textrm{sgn}(x)=-1$, $0$, and $1$ when $x<0$, $x=0$, and $x>0$, respectively. For simplicity, we consider only a hardcore repulsion interaction between the two donors.
Here, we adopt the thermal acceleration mechanism applied for SrTiO$_3$~\cite{Menzel}. So, high temperature $\beta \sim 15$ eV$^{-1}$ ($800$ K) can be used for our simulation with $U_0=1.01$ eV~\cite{Menzel}. From Eq.~(\ref{prob}) it is obvious that transformations as $\beta \rightarrow \beta/b$, $U_0 \rightarrow bU_0$, and $E \rightarrow bE$ do not change the hopping probabilities. Thus, we use $\beta = 40$ eV$^{-1}$ ($300$ K) and $U_0=0.379$ eV instead of $\beta = 15$ eV$^{-1}$ and $U_0=1.01$ eV. Here, the attempt frequency for the hopping is $10^{13}$ Hz~\cite{Jeon}. 

\begin{figure}[t]
\includegraphics[width=1.0\linewidth]{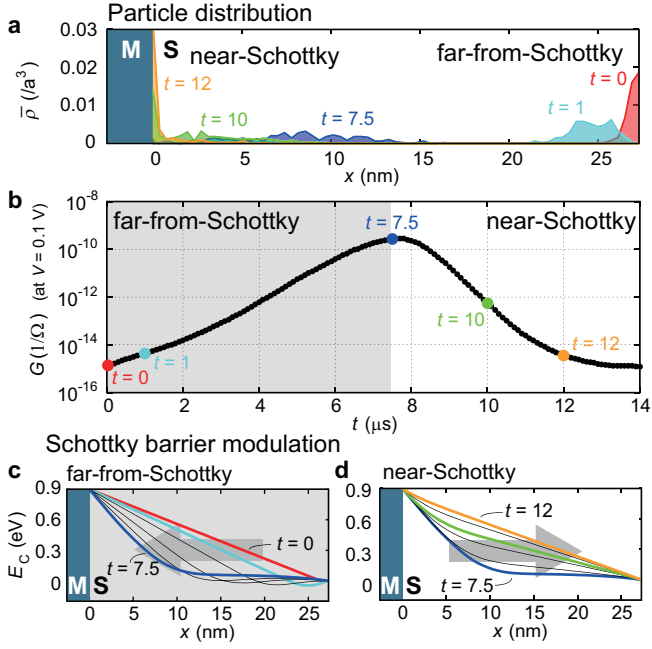}
\caption{(Color online) (a) Changes in the donor density distribution from $t=0$ to $14$ $\mu$s when a negative bias is applied. (b) Changes in the conductance as a function of time. (c) and (d) show changes of the Schottky barrier when most donors are distributed in the far-from-Schottky and near-Schottky regions, respectively. In all figures, red, cyan, blue, green, and gold are used to represent data collected at $t = 0$, $1$, $7.5$, $10$, and $12$ $\mu$s, respectively.} \label{result1}
\end{figure}

Using the above equations, the case in which the donors move from the Ohmic to the Schottky interface can be simulated. Initially, the donors were uniformly distributed with a density of $10^{19}~/\textrm{cm}^3$~\cite{Menzel}. Using Eq.~(\ref{prob}), we pushed the donors toward the Ohmic interface by applying a positive bias, the red curve in Fig.~\ref{result1}(a). Then, we applied a negative voltage $V_\textrm{ext}=-1.875$ V to attract donors towards the Schottky interface. Here, the donor density at $x_i$ is defined as $\overline{\rho}(x_i )\equiv n(x_i)/(L_y L_z)$, where $n(x_i )$ is the number of donors at the $x = x_i$ plane. Here, $\epsilon=100\epsilon_0$~\cite{Berg} in high electric field  ($\epsilon_0$ the permittivity in free space), with periodic boundary conditions in the $y$- and $z$-directions. Fig.~\ref{result1}(a) shows the time-dependent distribution of the donors. The distribution moved toward the Schottky interface over time.

The conductance $G(\equiv I/V_\textrm{ext})$ during the attraction process is calculated at $0.1$ V as a function of time $t$. As indicated in Fig.~\ref{result1}(b), the $G$-$t$ plot can be divided into two regions: for $t <7.5 ~\mu\textrm{s}$, $G$ increases as a function of $t$, and for $t > 7.5 ~\mu\textrm{s}$, $G$ decreases. When comparing the distributions shown in Fig.~\ref{result1}(a), $G$ increased (decreased) when most dopants were distributed in the far-from-Schottky (near-Schottky) region. 

The two different $t$-dependences of $R$ come from the different modulation behaviors of  the Schottky barrier during the attranction process. For each $t$, we obtained $E_C (x_i )$ by calculating $E_C (x_i,y_j,z_k )$ at $V_\textrm{ext}=0$ and averaging over $y_j$  and $z_k$. Fig.~\ref{result1}(c) shows $E_C (x_i )$ when most dopants were distributed in the far-from-Schottky region (i.e.,  $t < 7.5 ~\mu\textrm{s}$). In this case, the pulling of the donors toward the Schottky interface resulted in a decrease in the Schottky barrier width, and $G$ increased. Fig.~\ref{result1}(d) presents the case $E_C (x_i )$ where most dopants were distributed in the near-Schottky region (i.e.,  $t > 7.5 ~\mu\textrm{s}$). In this case, the attraction of the donor increased the Schottky barrier width rather than decreasing it. These results agree with those of the one-dimensional SMD model. 

\begin{figure}[t]
\includegraphics[width=1.0\linewidth]{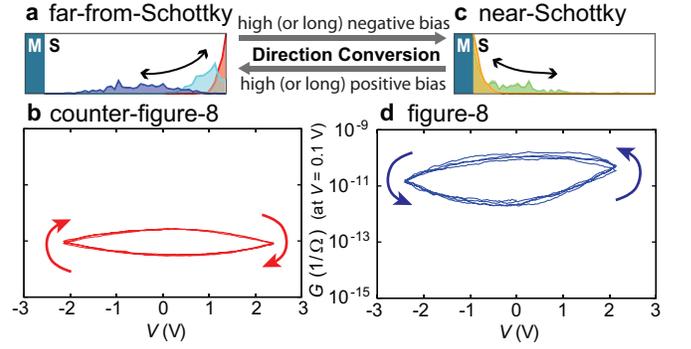}
\caption{(Color online) If the donors are initially distributed in the far-from-Schottky region (a), the voltage sweep results in a cF8 directional $G$-$V$ curve (b). If the donors are initially distributed in the near-Schottky region (c), a F8 direction (d) is obtained. By applying a large negative bias to the lattice exhibiting a F8 direction, we can attract donors into the near-Schottky region, and then the direction of $G$-$V$ curve will change to F8 way. The opposite effect can be obtained by applying a large positive bias.}\label{result2}
\end{figure}

We also simulated $G$-$V$ curves under a repetitive voltage sweep, with different initial donor distributions. Here, it took $0.1~\mu$s for each voltage point and voltage gap is $0.027$ V. During the voltage sweep, $G$ is calculated at $0.1$ V.
When most donors were initially distributed in the far-from-Schottky (near-Schottky) region as shown in Fig.~\ref{result2}(a) (Fig.~\ref{result2}(c)), $G$ decreases (increases) at positive bias and increases (decreases) at negative bias as shown in Fig.~\ref{result2}(b) (Fig.~\ref{result2}(d)), which corresponds to the direction of Fig.~\ref{analytic}(b) (Fig.~\ref{analytic}(c)) (see also movie clips). Furthermore, the direction of the $G$-$V$ curve can be changed by applying a large bias, as indicated in Fig.~\ref{result2}.


 
In conclusion, we introduced the SMD model which demonstrates that two opposite hysteresis curves intrinsically appear in the SMD due to the inhomogeneous dopant density distriubtion. From this theoretical analysis, we can control the type of the $I$-$V$ curve by modulating the mobile dopant distribution. The theoretical result we obtained in the Letter may become a fundamental basis for further development of SMD.

This research was supported by the National Research Foundation of Korea, Grants No. 2010-0015066 (B.K.), No. 2010-0020416 (T.W.N.), and No. NRF-2011-35B-C00014 (J.S.L.) and by the NAP of the Korean Research Council of Fundamental Science and Technology (B.K.).

\vfil\eject
\end{document}